\documentclass{aa} 
\usepackage[varg]{txfonts}
\usepackage{graphicx}
\usepackage{amssymb}    
\usepackage{color}
\usepackage{txfonts}
\usepackage{xcolor}
\usepackage{longtable}
\usepackage{siunitx} 
\usepackage{upgreek}
\usepackage{hyperref}
\usepackage[normalem]{ulem}
\usepackage{amstext}
\definecolor{ao(english)}{rgb}{0.0, 0.5, 0.0}

\newcommand{\presto}{{\sc presto}}
\newcommand{\sigproc}{{\sc sigproc}}

\begin{document} 

   \title{Searching for pulsars in the Galactic centre at 3 and 2 mm}

   \author{
   P.~Torne\inst{1,2}\thanks{Main contact author, \email{torne@iram.es} } \and
   G.~Desvignes\inst{3,2}\and
   R.~P.~Eatough\inst{4,2}\and
   M.~Kramer\inst{2,5}\and
   R.~Karuppusamy\inst{2}\and
   K.~Liu\inst{2}\and
   A.~Noutsos\inst{2}\and
   R.~Wharton\inst{2}\and
   C.~Kramer\inst{6,1}\and
   S.~Navarro\inst{1}\and
   G.~Paubert\inst{1}\and
   S.~Sanchez\inst{1}\and
   M.~Sanchez-Portal\inst{1}\and
   K.~F.~Schuster\inst{6}\and
   H.~Falcke\inst{7}\and
   L.~Rezzolla\inst{8,9}
   }

  \institute{
  Institut de Radioastronomie Millim\'etrique (IRAM), Avda. Divina Pastora 7, Local 20, 18012 Granada, Spain
  \and
  Max-Planck-Institut f\"{u}r Radioastronomie, Auf dem H\"{u}gel 69, D-53121, Bonn, Germany
  \and
  LESIA, Observatoire de Paris, Université PSL, CNRS, Sorbonne Université, Université de Paris, 5 place Jules Janssen, 92195 Meudon, France
  \and
  National Astronomical Observatories, Chinese Academy of Sciences, 20A Datun Road, Chaoyang District, Beijing 100101, PR China
  \and 
  Jodrell Bank Centre for Astrophysics, School of Physics and Astronomy, The University of Manchester, Manchester M13 9PL, UK
  \and
  Institut de Radioastronomie Millim\'etrique (IRAM), 300 rue de la Piscine, 38406 St. Martin d’Hères, France
  \and
  Department of Astrophysics/IMAPP, Radboud University Nijmegen P.O. Box 9010, 6500 GL Nijmegen, The Netherlands
  \and 
  Institut für Theoretische Physik, Max-von-Laue-Strasse 1, D-60438 Frankfurt, Germany
  \and
  Frankfurt Institute for Advanced Studies, Ruth-Moufang-Strasse 1, D-60438 Frankfurt, Germany
  }

  \date{Received Mar 10, 2021; accepted Mar 25, 2021}

 
  \abstract{
    Pulsars in the Galactic centre promise to enable unparalleled tests of gravity theories and black hole physics
    and to serve as probes of the stellar formation history and evolution and the interstellar medium in the complex central region of the Milky Way.
    The community has surveyed the innermost region of the galaxy for decades without detecting a population of pulsars, which is puzzling. A strong scattering of the pulsed signals 
    in this particular direction has been argued to be a potential reason for the non-detections.
    Scattering has a strong inverse dependence on observing frequency, therefore an effective way to alleviate its effect is to use higher frequencies in a survey for pulsars in the Galactic centre, in particular, close to the supermassive black hole Sagittarius~A*. 
    We present the first pulsar survey at short millimetre wavelengths, using several frequency bands between 84 and 156~GHz ($\uplambda\!=$3.57$-$1.92~mm), targeted to the Galactic centre. The observations were made with the Institut de Radioastronomie Millim\'etrique (IRAM)~30m Telescope in 28 epochs between 2016 December and 2018 May. This survey is the first that is essentially unaffected by scattering and therefore unbiased in population coverage, including fast-spinning pulsars that might be out of reach of lower-frequency Galactic centre surveys. We discovered no new pulsars and relate this result mainly to the decreased flux density of pulsars at high frequencies, combined with our current sensitivity. However, we demonstrate that surveys at these extremely high radio frequencies are capable of discovering new pulsars, analyse their sensitivity limits with respect to a simulated Galactic centre pulsar population, and discuss the main challenges and possible improvements for similar surveys in the future.
    }


   \titlerunning{Searching for pulsars in the Galactic centre at 3 and 2 mm}
   \authorrunning{P.~Torne et al.}

   \maketitle


\section{Introduction}




Pulsars in the Galactic centre of the Milky Way, in particular if found orbiting a stellar mass or intermediate-mass black hole, or the central supermassive black hole (SMBH) Sagittarius$\,$A* (hereafter Sgr$\,$A*), could allow direct accurate measurements of black hole properties such as mass, spin, or even quadrupole moment, and potentially enable the most stringent tests of general relativity and alternative theories of gravity to date \citep{wk99, kra04, pl04, liu12, psal16}.
In addition, Galactic centre pulsars would enable precision measurements of the interstellar medium (ISM) properties along the line of sight and at the central galactic region, such as electron density, scattering, or magnetic field \citep[e.g.][]{sj13, eat13, spi14, bow14, dex2017GCscatt}.
Moreover, discovering a population of pulsars in the nuclear cluster of the Milky Way could help us understand the enigmatic and complex star formation history and evolution in the region \citep[for a review, see e.g.][]{genzel10}.
Such pulsar-based measurements, in particular those related to Sgr$\,$A*, would furthermore strongly complement the ongoing efforts to understand black holes, their environments, and the theory of gravity through the study of orbital dynamics \citep{grav18, Do2019, grav20_precS2} or through event-horizon-scale imaging \citep{goddi17, eht19a, eht19f}.

Because of the scientific potential of pulsars located in the Galactic centre, significant efforts have been undertaken to try to find pulsars in the surroundings of Sgr$\,$A* \citep[e.g.][]{john95, kra2000, morr02, klein2005, john06, den09, mq10, bat11, whar12p1, eat13b, siem13}. Interestingly, all attempts to find a population of pulsars within the inner $\sim$20 arcmin of the Milky Way have been unsuccessful so far: only one radio magnetar, PSR$\,$J1745$-$2900, was found located $\approx$2.4 arcsec ($\approx$0.1 pc in projection) from Sgr$\,$A* \citep{ken13, mori13, rea13, eat13, sj13}. Unfortunately, PSR$\,$J1745$-$2900 is still too far from the SMBH to enable the desired gravity tests, although it allowed us to measure the gas properties close to Sgr$\,$A*, showing that the gas is dense, turbulent, highly magnetised, and varies on relatively short timescales \citep{eat13, des18}.

It is puzzling that no more pulsars are found in the central part of the Galaxy.
Massive stars populate the region with pulsars progenitors \citep[for a review, see e.g.][]{figer09}, and supernova remnants and pulsar wind nebulae candidates have been identified \citep{Wang2005, muno08, zhang2020}, as well as stellar mass black holes \citep{hailey18, gene18}. The scenario of a large population of Galactic centre pulsars is also supported by the Galactic centre excess of gamma-ray emission, which can be explained by an undetected large population of millisecond pulsars \citep[MSPs; e.g.][]{aba11, gormac13, bra15}. Theoretical estimates of the number of observable pulsars in the inner part of the Galaxy vary considerably between a few to several hundreds \citep{pl04, fau11, whar12, dexo14, chen14, raj18, schoedel20}, showing the complexity of the Galactic centre stellar population, evolution, and its environment.

Different theories trying to explain the lack of detections have been proposed. For example, a deficit of normal pulsars in the Galactic centre with an unusual population composed mainly of magnetars \citep[which rarely emit in radio, and whose emission may be intermittent and therefore difficult to detect;][]{dexo14}, or millisecond pulsars \citep[which might tend to be less luminous than normal pulsars and thus beyond the reach of our currently achievable sensitivities;][]{bai97, kra98, burg13, macq15}. Geodetic precession of pulsars due to massive companions could move their emission beams in and out our line of sight, making them undetectable during long periods of time \citep[e.g.][]{kra98, perera2010}, and eclipses may temporarily obscure the radio emission in certain binary systems \citep[e.g.][]{freire2005}. Similarly, for tight binary pulsars, certain epochs might face orbital ranges with characteristics that are difficult to detect by searching algorithms \citep[e.g.,][]{eat21sub}. Even dark matter has been considered, which could produce a rapid collapse of the neutron stars into black holes through accretion, making the life of pulsars unusually short in the inner Galactic centre region \citep{bra14}. 

Although some of the previously outlined reasons are plausible, the scattering of the pulsed signals by the dense turbulent gas in the direction of the Galactic centre is currently widely accepted as the most probable cause for the non-detections \citep{cl98, cl2002}. Nevertheless, \citet{spi14} and \citet{bow14} showed that the scattering in the direction of PSR$\,$J1745$-$2900, only $\approx$2.4 arcsec from Sgr$\,$A*, is much weaker than predicted by theoretical models \citep[see also][]{wuck14}.
A possible explanation for the low scattering measured for PSR$\,$J1745$-$2900 and the paucity of detections of other Galactic centre pulsars is that scattering in the Galactic centre region may be complex and strongly dependent on the line of sight because of multiple screens. This situation has also been found in other directions \citep[e.g.][]{roy2013, schni16, dex2017GCscatt}.

The ISM effects, and in particular, scattering, are strongly inverse dependent on the observing frequency \citep[see e.g.][]{lorkra04}. The detrimental effect of the scattering that may prevent the detection of the Galactic centre pulsars can therefore be significantly alleviated by increasing the frequency of the radio observations. The drawback of this approach is that pulsars, which usually are steep spectral sources \citep[$<\alpha>=-1.8\pm0.2$, for $S\propto\nu^{\alpha}$,][]{mar2000}, become extremely weak at high frequencies. However, some pulsars have shown flat spectral indices \citep[$\alpha>-1.0$, see e.g. the PSRCAT database,][]{man05} that keep the radio emission strong even at very high radio frequencies. Interestingly, certain pulsar emission models furthermore predict that incoherent emission may take over in pulsar spectra between radio and infrared wavelengths, potentially making the emission brighter at millimetre wavelengths than the power-law extrapolation from centimetre wavelengths \citep[see][]{mich78, kra96, kra97_7mm}. The highest radio frequency at which pulsations from a normal pulsar were detected is 138~GHz ($\uplambda$=2.17~mm) \citep[Torne et al. \emph{in prep.}]{tortesis17}, and radio magnetars have been detected up to $\sim$300$\,$GHz \citep{tor17, tor20}. This supports the hypothesis that we can try to discover pulsars using very high radio frequencies.

Because of the lack of pulsar discoveries and in order to overcome the ISM effects, in particular, the strong scattering, targeted surveys at the Galactic centre have progressively increased the radio frequency of the observations. The highest radio frequencies used to date for Galactic centre surveys were around 20~GHz \citep{siem13, eat13b}. In this work, we present a new targeted pulsar survey at the Galactic centre that for the first time uses short millimetre wavelengths between 3 and 2$\,$mm ($\nu \sim$ 84 to 156~GHz). The paper is structured as follows. The observations are introduced in Section ~\ref{sec:obs}, and the data analysis is explained in detail in Section~\ref{sec:dataanalysis}. Section
~\ref{sec:resdis} presents and discusses our results, and Section~\ref{sec:sumcon} includes a summary and our conclusions.

\section{Observations}\label{sec:obs}


The observations were made with the Institut de Radioastronomie Millim\'etrique (IRAM) 30m telescope on Pico Veleta, Spain, during different campaigns aimed primarily at monitoring the Galactic centre magnetar PSR$\,$J1745$-$2900 between December 2016 and May 2018 (under project numbers 159-16, 039-17, and 145-17). The receiver we used was the Superconductor-Insulator-Superconductor (SIS) Eight MIxer Receiver \citep[EMIR,][]{car12}, tuned to different frequencies in the 3.4, 2.1, 1.3, and 1.0~mm bands (from $\sim$80 to $\sim$300~GHz) at different epochs. For this pulsar survey, only the observations from the 3.4 and 2.1 mm bands (central frequencies 86 and 102, and 138 and 154$\,$GHz, hereafter referred to as 3 and 2 mm bands) are analysed. This selection was made because the telescope is more efficient at these bands\footnote{The main beam efficiencies are $\rm{B_{eff}}\simeq$ 80, 74, 55, and 40\% for the 3, 2, 1, and 0.8$\,$mm bands, respectively \citep{ckra13}. In addition, the typical system temperatures ($T_{\rm sys}$) are lower for the longer wavelength bands because of a lower atmospheric opacity \citep[see e.g.][]{pety09}.}, and pulsars should be brighter here than at even higher frequencies. Additionally, we minimise the effect on signal quality and sensitivity by atmospheric effects, in particular, the low-frequency noise and signal absorption, which are magnified by the low telescope elevations that are necessary to observe the Galactic centre from the IRAM~30m telescope (elevation $\lesssim$ 24 deg).

The backend used to record the signal from EMIR was the Broad-Band Continuum backend (BBC), which provided one time series with a resolution of 100$\,\mu$s for each frequency band and polarisation. This combination of EMIR and the BBC resulted in an effective non-contiguous bandwidth of $\sim$32$\,$GHz (four bands of $\sim$8~GHz centred at the above-mentioned central frequencies of 86, 102, 138, and 154$\,$GHz). With this set-up, each observation produced eight time series: from four frequency bands times two linear polarisations.

The observations at each epoch typically consisted of sessions of three to four hours, in which continuous tracks of PSR$\,$J1745$-$2900 of 10 to 20 minutes were alternated with calibration measurements. Tracking a source continuously is an observing strategy that differs from typical observing methods in millimetre astronomy, where the beam is usually switched on and off the target source to remove undesired instrumental and atmospheric effects. The reason for using tracking when observing pulsars is mainly twofold: to keep the coherence of the time series, which is not ensured if the telescope uses beam-switching techniques; and to gain sensitivity because with beam-switching techniques half of the time is spent off source (losing $\rm \sqrt{2}$ of sensitivity), with an additional $\rm\sqrt{2}$ of loss when the off is subtracted.

The beam size of the IRAM~30m telescope is $\approx\,$29, 24, 18, and 16~arcsec at 86, 102, 138, and 154~GHz, respectively. Sgr$\,$A* is 2.4$\pm$0.3 arcsec from PSR$\,$J1745$-$2900 \citep{rea13}, therefore it is included close to the centre of the beams for all frequencies. In the worst case, that is, for the smallest beam sizes at the 2 mm band, the 2.4 arcsec separation from Sgr A* would reduce the sensitivity by $\lesssim$20\%, reaching negligible reductions of sensitivity for the 3 mm beams. At a distance to the Galactic centre of $d_{\rm GC}=\,$8.18~kpc \citep{gravc19}, the projected diameter around PSR$\,$J1745$-$2900 is 1.16, 0.96, 0.72, and 0.64~pc for each of the four central observing frequencies.

The total number of analysed epochs between December 2016 and May 2018 is 28. The observations amount to 62.2 hr covering the Galactic centre. The individual epochs are presented in Table~\ref{tab:obs_table}.

\begin{table*}
  \centering
  \caption{List of the observations of the Galactic centre analysed in this survey. Columns show the date, the modified Julian Date of the start of each observation ($\rm MJD_{start}$), the integration time on-source (${t_{\rm int}^{\rm GC}}$), the total time of each searched time series after concatenating and padding gaps (${t_{\rm int}^{\rm GC+GAPS}}$, see Sec.~\ref{sec:dataanalysis_dcp}), the average system temperature of the different datsets (3, 2, and 3+2~mm) on the line of sight ($\left<T_{\rm sys}\right>$), the number of candidates produced per searched epoch ($\rm{N_{cands}}$), and the IRAM project number to which the observation corresponds.}
  \label{tab:obs_table}
   \begin{tabular}{c@{\hskip 0.6cm}c@{\hskip 0.6cm}c@{\hskip 0.45cm}c@{\hskip 0.45cm}c@{\hskip 0.45cm}c@{\hskip 0.45cm}c@{\hskip 0.45cm}c@{\hskip 0.45cm}c@{\hskip 0.45cm}c@{\hskip 0.45cm}c@{\hskip 0.45cm}}
\hline
   Date         & MJD$_{\rm start}$             & ${t_{\rm int}^{\rm GC}}$      & ${t_{\rm int}^{\rm GC+GAPS}}$  &        $\left<T_{\rm sys} ^{\rm 3mm}\right>$   &       $\left<T_{\rm sys}^{\rm 2mm}\right>$  &    $\left<T_{\rm sys}^{\rm 3+2mm}\right>$ &   $\rm{N_{cands}^{3mm}}$  &       $\rm{N_{cands}^{2mm}}$  &       $\rm{N_{cands}^{3+2mm}}$ & Project\\
                                &                               &   (hr)    & (hr)  & (K)   &   (K)   &  (K)  &        &      &        &\\
\hline
   2016 Dec 12  &   57734.459837    & 2.0  &   3.9 & 118.2   & 126.6  & 122.4  &   141 &   25  &   76   & 159-16\\
   2016 Dec 21  &   57743.406261    & 2.2  &   3.2 & 117.7   & 158.4  & 138.0  &   130 &   22  &   116  & 159-16\\
   2016 Dec 22  &   57744.399016    & 1.8  &   3.3 & 141.5   & 225.2  & 183.3  &   148 &   56  &   120  & 159-16\\
   2017 Jan 02  &   57755.366388    & 1.8  &   3.1 & 137.8   & 213.5  & 175.6  &   55  &   46  &   67   & 159-16\\
   2017 Jan 23  &   57776.368090    & 2.5  &   3.0 & 113.1   & 137.4  & 125.3  &   70  &   70  &   184  & 159-16\\
   2017 Jan 30  &   57783.304502    & 2.4  &   3.2 & 146.1   & 225.2  & 185.7  &   89  &   89  &   123  & 159-16\\
   2017 Feb 14  &   57798.268530    & 2.3  &   3.0 & 110.0   & 130.9  & 120.4  &   96  &   61  &   105  & 159-16\\
   2017 Mar 01  &   57813.330613    & 1.0  &   1.1 & 110.7   & 119.9  & 115.3  &   244 &   247 &   236  & 159-16\\
   2017 Mar 10  &   57822.205694    & 2.5  &   3.0 & 107.4   & 123.8  & 115.6  &   37  &   28  &   49   & 159-16\\
   2017 Mar 28  &   57840.163564    & 3.8  &   4.6 & 126.1   & 152.2  & 139.1  &   15  &   45  &   36   & 159-16\\
   2017 Apr 18  &   57861.084317    & 2.0  &   2.4 & 119.9   & 155.3  & 137.6  &   119 &   68  &   134  & 159-16\\
   2017 May 18  &   57890.996898    & 1.7  &   3.1 & 145.1   & 234.5  & 189.8  &   12  &   26  &   30   & 159-16\\
   2017 May 24  &   57896.989999    & 1.3  &   2.2 & 164.3   & 303.3  & 233.8  &   38  &   42  &   49   & 159-16\\
   2017 Jun 05  &   57909.951793    & 2.1  &   4.1 & 153.1   & 262.9  & 208.0  &   11  &   12  &   9    & 039-17\\   
   2017 Jul 03  &   57937.904224    & 1.3  &   3.2 & 115.7   & 140.4  & 128.0  &   44  &   42  &   49   & 039-17\\
   2017 Sep 04  &   58000.726469    & 3.0  &   4.4 & 139.3   & 209.9  & 174.6  &   81  &   69  &   76   & 039-17\\
   2017 Sep 05  &   58001.711724    & 3.7  &   4.9 & 173.4   & 310.7  & 242.1  &   25  &   20  &   28   & 039-17\\
   2017 Sep 28  &   58024.738425    & 1.7  &   2.0 & 222.0   & 504.2  & 363.1  &   49  &   46  &   64   & 039-17\\
   2017 Oct 09  &   58035.638182    & 2.7  &   4.1 & 161.5   & 274.7  & 218.1  &   29  &   42  &   43   & 039-17\\   
   2017 Nov 08  &   58065.535856    & 2.7  &   4.1 & 113.0   & 134.3  & 123.6  &   30  &   32  &   32   & 039-17\\
   2017 Nov 20  &   58077.513564    & 2.5  &   3.5 & 123.5   & 164.1  & 143.8  &   23  &   31  &   25   & 039-17\\
   2017 Dec 04  &   58091.491030    & 2.3  &   2.9 & 113.7   & 144.5  & 129.1  &   51  &   51  &   68   & 145-17\\
   2017 Dec 18  &   58105.442222    & 1.7  &   2.7 & 111.3   & 140.2  & 125.8  &   37  &   79  &   98   & 145-17\\
   2018 Jan 03  &   58121.407222    & 3.0  &   3.7 & 116.5   & 148.5  & 132.5  &   38  &   79  &   157  & 145-17\\
   2018 Jan 15  &   58133.380520    & 2.3  &   3.3 & 121.2   & 146.9  & 134.0  &   41  &   38  &   52   & 145-17\\
   2018 May 01  &   58239.095451    & 2.0  &   3.6 & 110.3   & 116.2  & 113.2   &   42  &   16  &   101 & 145-17\\
   2018 May 15  &   58253.030335    & 2.1  &   3.2 & 118.5   & 147.1  & 132.8   &   27  &   11  &   18  & 145-17\\
   2018 May 22  &   58259.986226    & 1.8  &   3.1 & 147.5   & 239.8  & 193.7   &   28  &   12  &   20  & 145-17\\
\hline
\end{tabular}
\end{table*}

\section{Data analysis}\label{sec:dataanalysis}

The data analysis was split into four parts: calibration, data cleaning and preparation, pulsar searching, and pre-processing tests with synthetic pulsar signal injection and sensitivity analysis. We describe each of them in the following subsections.

\subsection{Calibration}\label{sec:dataanalysis_cal}

The first step of the analysis consisted of calibrating the data by calculating conversion factors from counts to Jansky and multiplying the time series correspondingly. This was made individually for each frequency band and each polarisation channel. The values required to calculate the receiver temperature and telescope gain were computed from a hot-cold calibration on loads of known physical temperature. The opacity of the atmosphere was calculated through a short measurement on sky 600 arcsec away in right ascension from the science target, using the information from the telescope weather station and the model called atmopsheric transmission at microwaves (ATM) \citep{pardo07} to derive the equivalent sky temperature. With this, we calculated the calibration temperature, $T_{\rm cal}$, which was used to convert from counts into ${T_{\rm a}}^*$, that is, antenna temperature outside the atmosphere \citep[see][]{ckra97}. The scaling factor from ${T_{\rm a}}^*$ to Jansky scale, $S/{T_{\rm a}}^*$, is obtained from the observatory efficiency tables\footnote{\url{https://www.iram.es/IRAMES/mainWiki/Iram30mEfficiencies}}. The $S/{T_{\rm a}}^*$ factor is not given for all frequencies, therefore a second-order polynomial was fit to those provided, assuming an error of 10 and 15\% for the 3 and 2 mm bands, respectively. We obtain $S/{T_{\rm a}}^* =$ 5.9, 6.0, 6.3, and 6.5 Jy/K for 86, 102, 138, and 154~GHz. Finally, a correction to the gain due to gain dependence on the telescope elevation was applied \citep{pen12}. The calibration measurements were made every 10 to 20 minutes to follow possible gain variations and opacity changes through observations. The final calibration factor applied to each time series was calculated for the middle time of each observation by interpolating linearly between the calibration results just before and after each scan.

\subsection{Data cleaning and preparation}\label{sec:dataanalysis_dcp}

We used a receiver designed for spectroscopic observations, for which long-term stability was not a requirement. In our fast-sampled long integrations,\footnote{For reference, spectroscopic scans with EMIR typically integrate for about 30 seconds and with sampling times longer than $\sim$100$\,$ms.} the data therefore occasionally showed short period of instabilities with an increased noise amplitude towards negative values and narrow drop-offs of the signal intensity. In addition, the data showed an excess of power weighted towards the low-frequency part of the Fourier spectrum (hereafter referred to as red noise). Red noise mainly arises from the atmospheric opacity variations during the observations. After the data were calibrated, we therefore applied a filtering to subtract or reduce the instrumental effects and the red noise. 

In a first step, the intensity drop-offs were clipped by a moving a window of 3.5$\,$s over the time series, substituting the samples that were below $-$5$\sigma$ with the median value of the samples within the window. The drop-offs occurred only a few times per scan, and their filtering does not affect our sensitivity. This running-clip filtering also partly reduced the short-term instabilities that show increased negative amplitudes in the noise when they exceed $-$5$\sigma$. Next, a Fourier transform was carried out, and a number of frequency bins were zapped (i.e. the real and imaginary parts were set to the average value and zero, respectively) to filter a strong 1 Hz signal caused by the cryo generator cycle and the 50 Hz power mains. Five and nine harmonics of the two signals  were also zapped. We then returned to the time domain by applying an inverse Fourier transform and applied a filtering to reduce the red noise, which consists of a moving window of 3$\,$s that subtracts a fitted polynomial of first order. This resulted in time series that were flat and did almost not affect the potential pulsations, whether inside the noise or even potential single pulses. However, we note that this filtering scheme may cause very slowly rotating pulsars to appear below the noise mean level after folding. Our tests showed that only periods above $P\gtrsim$4~s may be significantly affected, which reduces our sensitivity to long-period pulsars both in the periodicity search and when folding to produce candidate plots. Finally, we also note that the noise from the EMIR receiver shows locally generated periodic signals of strong intensity that proved difficult to model and filter because their amplitude and frequency structures vary slightly in time and can in addition vary with each new receiver tuning. We used an adaptive filtering in the Fourier domain to partly subtract these signals and discuss its potential effect in our pulsar detection capabilities in Sec.~\ref{sec:resdis}.


After the calibration and the cleaning, the two polarisation channels of each frequency band were summed, producing the total intensity time series per frequency band. At this stage, we had four time series per scan (one for each frequency band). From these total intensity time series, three different datasets were created: One set combines the two frequency bands at 3 mm (86+102~GHz), one set combines the two frequency bands at 2 mm (138+154~GHz), and a third set combines the four frequency bands (86+102+138+154~GHz), which we name 3+2 mm. The combination was made by averaging together the total intensity time series of the different frequency bands. We remark that the time series of the different frequencies are aligned in time, that is, the start time for all of them is exactly the same, thus enabling this simple averaging. 
We discuss the reasons for not focusing on the theoretically most sensitive 3+2$\,$mm dataset alone, which includes the effect of uncorrected interstellar dispersion and technical motivations, in Sec.~\ref{sec:resdis}.

Lastly, as the observations were split into scans of 10$-$20~min, we coherently concatenated together all the scans for each dataset per epoch. The concatenation was made by filling the gaps with the median value (which was zero after the red noise filtering applied before). The total length of the concatenated time series varied for different epochs, with typical values around 3 to 4~hr. 

\begin{figure*}
    \centering
    \includegraphics[width=\textwidth]{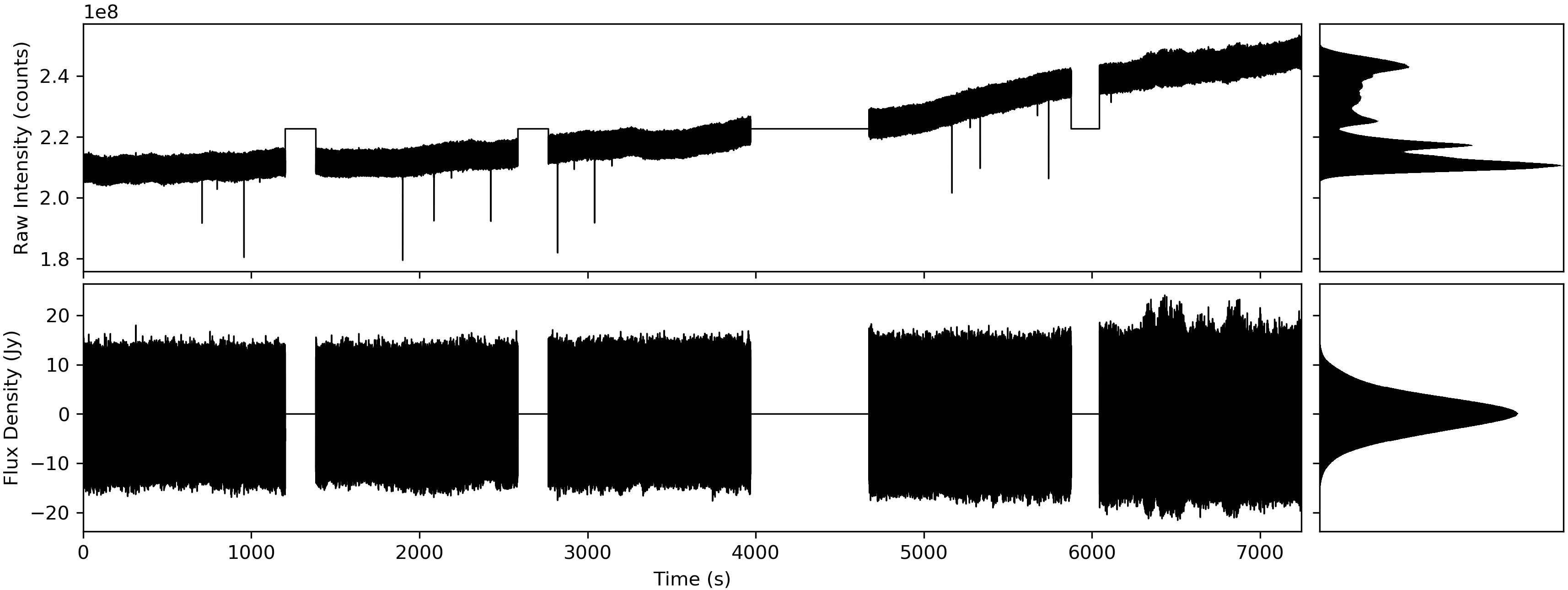}
    \caption{Visualisation of an example time series of the data from EMIR plus BBC before (upper panels) and after (bottom panels) the processing presented in Sections~\ref{sec:dataanalysis_cal}, \ref{sec:dataanalysis_dcp}. The processing includes calibration, data cleaning, and preparation prior to the pulsar searching. The upper left panel shows the raw data as recorded by the BBC during an observation consisting of five scans of 20$\,$min each during a total time span of about 2$\,$hr. The atmospheric opacity variations and the sporadic intensity drop-offs are apparent. The gaps in between scans show intervals when the telescope was not observing the Galactic centre, e.g., in the course of calibration scans, or pointing and focus adjustments. The upper right panel shows a histogram with the distribution of the count levels through the observation, showing that it is not ideal to use the data directly in this state for statistics-based algorithms such as the pulsar searching. The bottom left and bottom right panels show the same time series after the processing. The slow level variations and the intensity drop-offs are corrected for. Furthermore, the intensity scale at this stage is calibrated in Jansky, which is necessary in order to correctly combine the different polarisations and frequency bands. The bottom left panel also shows the slight increase in the standard deviation of the noise toward the end of the observation. This is because the Galactic centre decreases in elevation on sky, which forces the telescope to observe through a larger amount of atmosphere (or airmass), thus increasing the system temperature. The data shown correspond to the 102~GHz frequency band from 2017 September 28. This epoch was chosen to illustrate a case with a high opacity and thus a high system temperature, which typically is more challenging to clean.} \label{fig:data_example}
\end{figure*}

Figure~\ref{fig:data_example} shows an example of data before and after the calibration, data cleaning, and preparation prior to searching.
Table~\ref{tab:obs_table} summarises the observations and presents the data length before and after concatenation of scans, and the average system temperature per epoch and frequency band.

\subsection{Pulsar searching}\label{sec:dataanalysis_ps}

We performed the search with a pipeline based on the {\sc presto} software version 2.1\footnote{\url{https://github.com/scottransom/presto}} \citep{ransom2001}. We searched the three datasets (3, 2 and 3+2~mm) of each epoch individually. The pipeline consists of the following searching algorithm.

\begin{figure*}
    \centering
    \includegraphics[width=\textwidth]{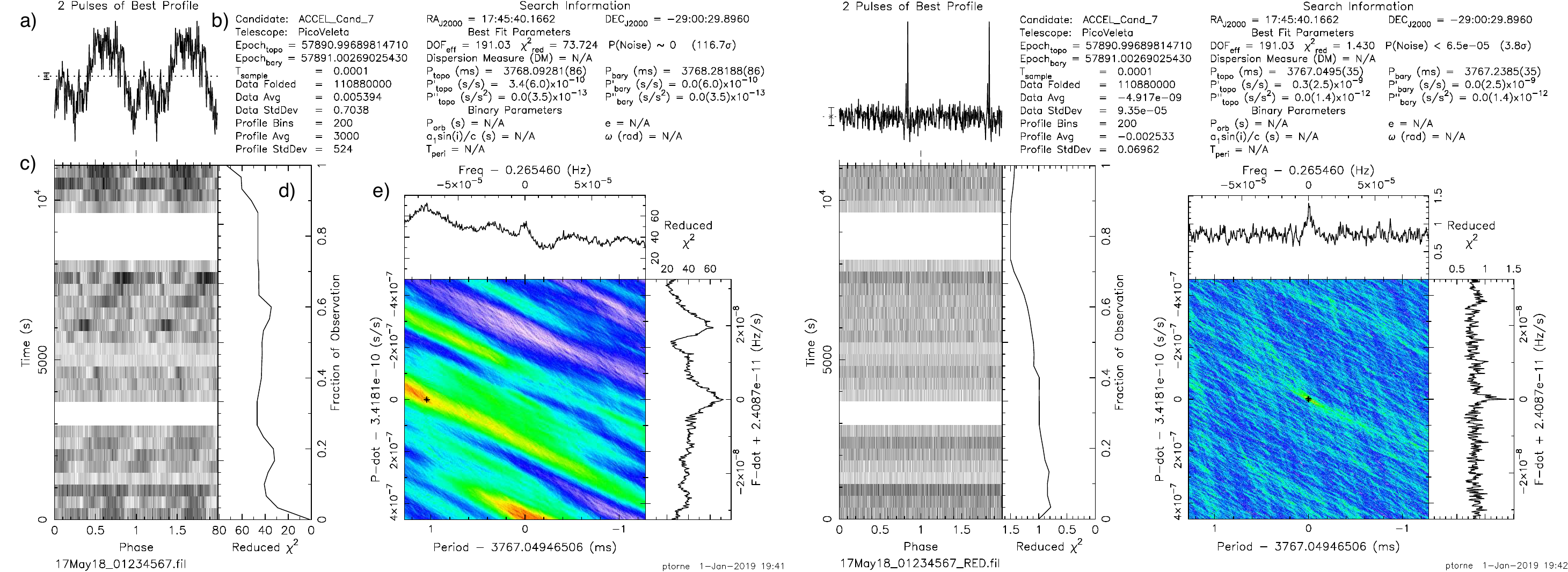}
    \caption{Example of the ability and importance of the \presto\  \texttt{rednoise} filter to reduce the EMIR locally generated signals that interfere with the search algorithm. The plots show a blind detection of the Galactic centre magnetar PSR$\,$J1745$-$2900 in one epoch in which the pulsar was relatively weak (2017 May 18, $S_{\rm J1745}^{3+2mm}=1.52\pm0.06\,$mJy). The left panel shows the candidate plot from \texttt{prepfold} when the raw data are folded and \texttt{rednoise} is not applied. The right plot shows the same candidate when the data are cleaned by \texttt{rednoise} prior to folding. The filter does subtract part of the pulsar signal, as apparent by the narrow profile and the dip in the profile below the mean noise level, but the improvement compared to a non-filtered version is so substantial that the pulsar may easily have been missed if the filtering had not been applied. The main panels in each plot show a) the folded profile twice, b) the candidate and observation details including name, telescope, epoch, time sampling, statistics, source coordinates, period and period derivatives (in topocentric and barycentric references), and binary parameters if applicable, c) waterfall plot of folded signal intensity vs. time, d) accumulated reduced $\upchi^2$ of the integrated profile vs. a model of noise alone, and e) two-dimensional waterfall plot of the reduced $\upchi^2$ of the candidate profile vs. noise as a function of folding period and period derivative. The white gaps in the phase vs. time panel are intervals in which the telescope did not observe the source, e.g. during calibration, pointing, or focus, or for some epochs, when observations were taking place at different frequency bands that were not used for the pulsar search (see Sec.~\ref{sec:obs}).} \label{fig:J1745_rednoise}
\end{figure*}

In a first step, we analyse the data for interfering signals using \texttt{rfifind}, which tries to identify chunks of bad data by comparing global and block-size statistics. \texttt{rfifind} produces a file with a mask that can later be used to flag specific blocks of data identified as having poor statistics. This automatic analysis did not always work; it often masked 100\% of the data. We attribute this large masking to the somewhat peculiar statistics of the noise from EMIR, which can deviate from well-behaved Gaussian noise, especially when integrated without off-source-position subtraction (see Section~\ref{sec:obs}). Whenever the masked fraction was above 30\%, we passed a filtered version of the data  to \texttt{rfifind} that we created using the \texttt{rednoise} filter of {\sc presto} on each scan before concatenating them. The \texttt{rednoise} filter acts in the Fourier domain by subtracting the slope of the power series using adaptive window sizes. A normalisation is also applied by \texttt{rednoise} so that the standard deviation of the powers in the filtered series equals one. This filtering proved very effective in reducing the undesired periodic instrumental signals in the data and improving the data statistics (see Fig.~\ref{fig:J1745_rednoise}). However, \texttt{rednoise} can also partially filter potential pulsar signals when their spin frequency is very low or very high. For this reason, we did not use it to remove the red noise during data preparation, and opted for the running-fit filter (see Sec.~\ref{sec:dataanalysis_dcp}). The program \texttt{rfifind} worked better in this \texttt{rednoise}-filtered version of the data, although the analysis of certain epochs still resulted in unacceptably high masking. If the masking fraction was still above 30\% after the \texttt{rednoise} filter was applied, we did not allow \texttt{rfifind} to automatically compute a bad-data mask. The typical percentage of the automatically masked fraction of data after this routine is about 1$-$3\%. After these steps, we manually passed the intervals of the gaps between scans that where
filled with the median value to \texttt{rfifind}  so that they were excluded during data analysis.

A barycentered time series was then produced with \texttt{prepsubband} using the previously created \texttt{rfifind} mask to substitute bad data blocks with median values, and without dispersion correction. The time series was then Fourier transformed, the \texttt{rednoise} filter applied to the Fourier series, and a number of persistent periodic instrumental signals from a database created in pre-processing tests were removed by zapping their corresponding frequency bins. The resulting Fourier series was searched for pulsar candidates with \texttt{accelsearch}, a routine that detects significant periodicities in the Fourier domain, including harmonic summing and a template-matched algorithm to recover signals with a Doppler shift (as occurs in pulsars that are accelerated when they orbit companions).

Two passes of \texttt{accelsearch} were made. The first pass set the \presto\  \texttt{zmax} parameter to zero. This \texttt{zmax} parameter is the maximum width in Fourier bins of the templates that are used to recover accelerated signals in the Fourier domain \citep{ranein2002}. Thus, the first pass is sensitive to isolated or very lowly accelerated pulsars. In order to be sensitive to potential binary systems, including pulsars orbiting stellar and intermediate-mass black holes and the central SMBH, the \texttt{zmax} parameter of \texttt{accelsearch} was set to 1200 in a second pass. This is the maximum currently allowed by the software. This high \texttt{zmax} parameter is motivated by the requirement that as many harmonics as possible need to be recovered to increase the sensitivity, in combination with the long total integration times, in particular after concatenating the scans per epoch (see~Table~\ref{tab:obs_table}). We remark that the drift in the Fourier domain ($\dot{r}$) due to the frequency shift ($\dot{f}$) by the Doppler effect of an accelerated pulsar has a quadratic dependence on integration time ($t_{\rm int}$), $\dot{r}=\dot{f}({t_{\rm int}})^2$ \citep{ransom2001}. In both passes we allowed for summing up to 32 harmonics, and set the sigma threshold to 2.0 and 3.0 for the \texttt{zmax}~=~0 and 1200, respectively. The candidates from the two passes were then sifted to harmonically remove related candidates, duplicates, and candidates with sifting significance < 2$\sigma$. In addition, we only required candidates to be detected with one harmonic to pass the sifting filter. Although the sigma thresholds may look low, pre-processing tests conducted to find optimum parameters for the pipeline found that setting the significance limits this low enabled detections of weak pulsar signals in the data (see Sec.~\ref{sec:fakepsr}) while still producing a manageable number of total candidates. The number of candidates produced in each epoch is shown in Table~\ref{tab:obs_table}. Typically, there were well below 200 candidates per observation.

The last step in the pipeline folds the data and produces plots to visually analyse the final candidates. This was done with \texttt{prepfold}. We carried out the folding twice for each candidate. At first, we folded the concatenated filterbank that is passed to the pipeline. Then we folded a concatenated filterbank that is filtered by the \texttt{rednoise} filter with the same parameters. The reason for folding twice is that the \texttt{rednoise}-filtered data tend to show a significantly cleaner profile and candidate signal (i.e. it is effective in removing parts of the local periodic interfering signals, as mentioned earlier), but the filtering may also subtract part of the candidate signal (see~Fig.~\ref{fig:J1745_rednoise}). Thus, by producing and reviewing the two folded candidate plots (original and filtered), we avoid missing a potential real pulsar by a filtering that is too strong or insufficient.

\begin{figure}
    \centering
    \includegraphics[width=\linewidth]{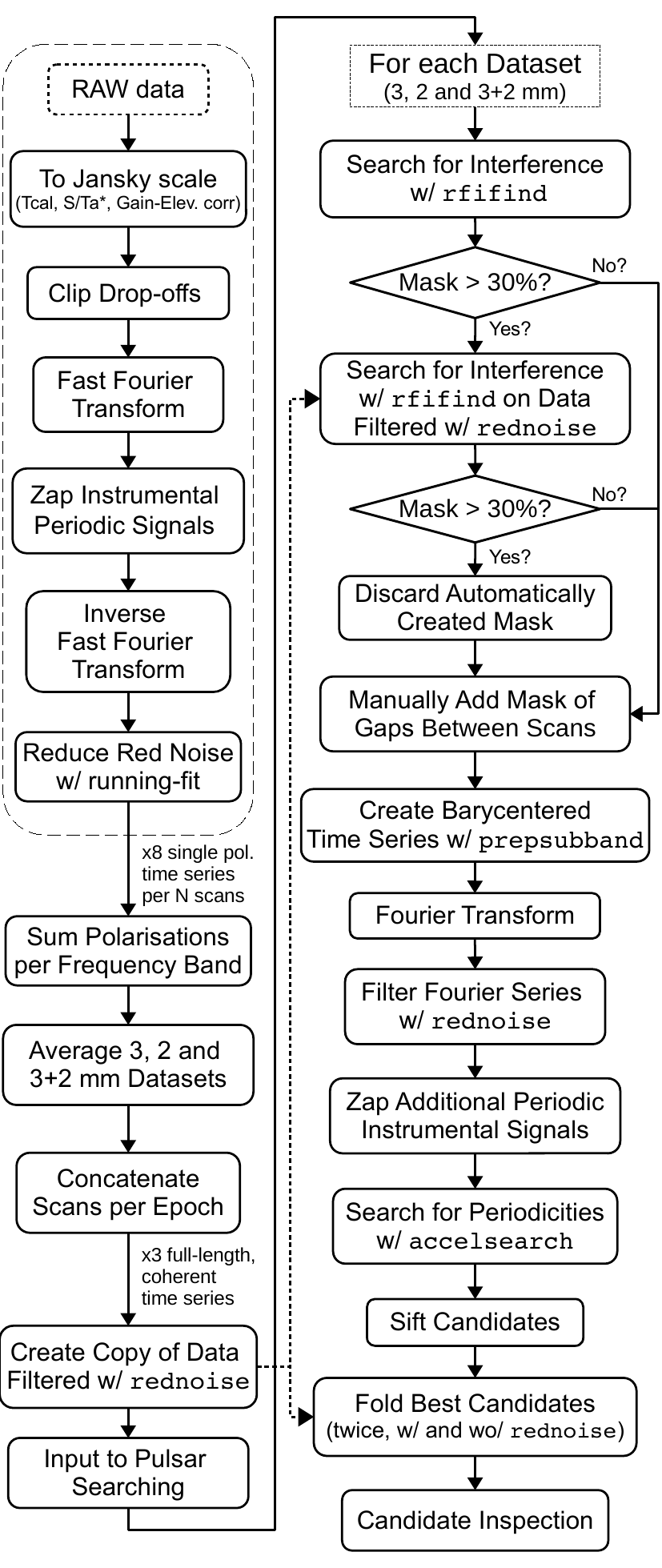}
    \caption{Flowchart of the data analysis as described in Section~\ref{sec:dataanalysis}. The algorithm is applied to each observing epoch individually. The steps inside the dashed block are applied in parallel to each single-polarisation time series of each frequency band. When this block finalises, the algorithm continues to reduce the data, creating total intensity time series and averaging frequency bands to create the three datasets 3, 2, and 3+2~mm (see Sec.~\ref{sec:dataanalysis_dcp}). These datasets then enter the searching algorithm as summarised in the right column of the diagram.} \label{fig:flowchart}
\end{figure}

Finally, we selected a number of observations made under good weather conditions, at different epochs, and searched them with a modified version of the pipeline that included searching in jerk space (i.e. derivative in acceleration). Everything was analogous in these cases except for the parameters for the acceleration search of \texttt{accelsearch}, which were \texttt{zmax}=300 and \texttt{wmax}=900. The parameter \texttt{wmax} controls the maximum size of the matched templates in the $\ddot{f}$ dimension \citep[][]{and18}. A visual summary of the data analysis is shown in Figure~\ref{fig:flowchart}.

\subsection{Sensitivity analysis}\label{sec:sensitivity}

To estimate the survey sensitivity, we used the radiometer equation adapted for pulsar observations to calculate a minimum detectable flux density \cite[see e.g.][]{lorkra04},

\begin{equation}\label{eq:radiometer}
S_{\rm min} = \frac{{\rm (S/N)_{min}} \, T_{\rm sys}}{\rm G \, \sqrt{\rm n_{p} \, t_{\rm int} \, \Delta \nu}} \; \sqrt{\frac{W_{\rm eff}}{P-W_{\rm eff}}},
\end{equation}

\noindent where $\rm (S/N)_{min}$ is the required signal-to-noise ratio for a detection, $T_{\rm sys}$ is the telescope system temperature, G is the telescope gain\footnote{\url{https://www.iram.es/IRAMES/mainWiki/Iram30mEfficiencies}}, $\rm n_{p}$ is the number of polarisations, $t_{\rm int}$ is the integration time, $\Delta \nu$ is the instantaneous observing bandwidth, and $W_{\rm eff}$ and $P$ are the effective width of the pulse and spin period of the pulsar, respectively. Taking into account the effects of scattering, dispersion, and instrumentation on the pulse widths, $W_{\rm eff}$ is calculated as the sum in quadrature of the intrinsic width of the pulse profile $W_{\rm int}$, the characteristic scattering time $\tau_{\rm s}$, the dispersion smearing across one frequency channel $\Delta t_{\rm DM}$, and the time resolution of the digitised data $\delta t$ \citep[e.g.,][]{man96},

\begin{equation}\label{eq:weff}
    W_{\rm eff} = \sqrt{{ W_{\rm int}}^2 + {\tau_{\rm s}}^2 + {\Delta t_{\rm DM}}^2 + {\delta t}^2}.
\end{equation}

Given the relatively varied dataset presented in this work, with observations taken at different frequency bands and weather conditions (and therefore with different $T_{\rm sys}$) and of different total integration time, we selected representative values of observations of good quality for our sensitivity analysis, that is, those obtained when the integration time was long and the weather was good. We focus on the scenario with the averaged 3+2~mm dataset (i.e. a central observing frequency of 120$\,$GHz with a bandwidth of 32$\,$GHz) as it is considered the most sensitive due to the large bandwidth. 

Inside Eq.~\ref{eq:radiometer}, two parameters depend on the particular pulsar observed: the spin period $P$, and the pulse width, $W_{\rm eff}$. These two parameters can be expressed combined in terms of duty cycle, $\delta=W_{\rm eff}/P$. Typical duty cycles are about 5 and 20\%  for normal and millisecond pulsars, respectively \citep[see][and references therein]{kra98}. We assumed a main scenario with $\delta=0.1$, that is, a duty cycle of 10\%. For completeness and to illustrate how the duty cycle and uncorrected smearing by dispersion affect the results, we discuss scenarios with $\delta=0.05$ and $\delta=0.4$ and include the effect of the uncorrected smearing in Sec.~\ref{sec:resdis}. For comparison and to place the sensitivity of the IRAM~30m telescope into context, we considered a potential and realistic survey with the Atacama Large Millimeter Array (ALMA)\footnote{ALMA offers from Cycle~8 a pulsar observing mode through its phasing system \citep{app18}. Phased-ALMA can be equivalent up to a $\sim$73 m dish \citep{alma_tr8}. We note that under very good weather and array performance, phased-ALMA system temperature and gain at Band 3 can improve to $T_{\rm sys}\simeq55\,$K and $\rm{G}\simeq1.15\,$K/Jy, respectively \citep{liu21sub}. These values are derived from the ALMA Level 2 Quality Assurance (QA2) calibration tables (available in the ALMA archive) for experiments 2016.1.00413.V and 2017.1.00797.V. following the procedures outlined in \citet{goddi19}. We chose the more general and constraining values from \citet{alma_tr8} for our sensitivity calculations.}. As ALMA has a better visibility of the Galactic centre, we assumed a total on-source integration time of 5 hr, with the maximum bandwidth set to 8~GHz, limited by the current capability of the recording backends \citep{alma_tr8}. Table~\ref{tab:readiometer} summarises the parameters we used to calculate the minimum detectable flux densities, resulting in $S_{\rm min}^{\rm 30m} \simeq 0.059\,$mJy and $S_{\rm min}^{\rm ALMA} \simeq 0.008\,$mJy for the IRAM~30m and for ALMA, respectively.

\begin{table}
        \centering
        \caption{Parameters used to calculate the minimum detectable flux density of the IRAM~30m survey, and for comparison, a potential Galactic centre pulsar survey with phased ALMA. Columns indicate the system temperature ($T_{\rm sys}$), the telescope gain (G), the number of polarisations (${\rm n_{p}}$), the on-source integration time ($t_{\rm int}$), the instantaneous bandwidth ($\Delta \nu$), and the duty cycle of the pulsar to be detected ($\delta$). The values are chosen to represent a sensitive observation at a frequency of 120~GHz. In both cases the signal-to-noise ratio required for the detection, $\rm (S/N)_{min}$, is set to 6.}
        \label{tab:readiometer}
        \begin{tabular}{cccccccc} 
                \hline
        &  $T_{\rm sys}$                & G &   ${\rm n_{p}}$   &       $t_{\rm int}$   &       $\Delta \nu$    &       $\delta$    &    $S_{\rm min}$
        \\
            &  (K)   & (K/Jy) &                 &       (hr)    &       (GHz)   &   & (mJy)
        \\
                \hline
        30m &  125              & 0.16 &   2   &   3.0 &   32  &   0.1 &   0.059 \\
        ALMA    & 70            & 1.05 &   2   &   5.0 &   8  &   0.1  &  0.008   \\
                \hline
        \end{tabular}
\end{table}

The $S_{\rm min}$ value can be used to directly compare the sensitivity with those of other surveys, but to better estimate the potential of this survey of detecting pulsars in the Galactic centre, we simulated a scenario in which we populated the Galactic centre region that is covered by our telescope beam with pulsars resembling those that are known to exist in the Milky Way. Then we calculated the percentage of these presumed Galactic centre pulsars that emit above the minimum detectable threshold. Those pulsars emitting above the threshold could have been discovered by the survey.

\begin{figure}
    \centering
    \includegraphics[width=\linewidth]{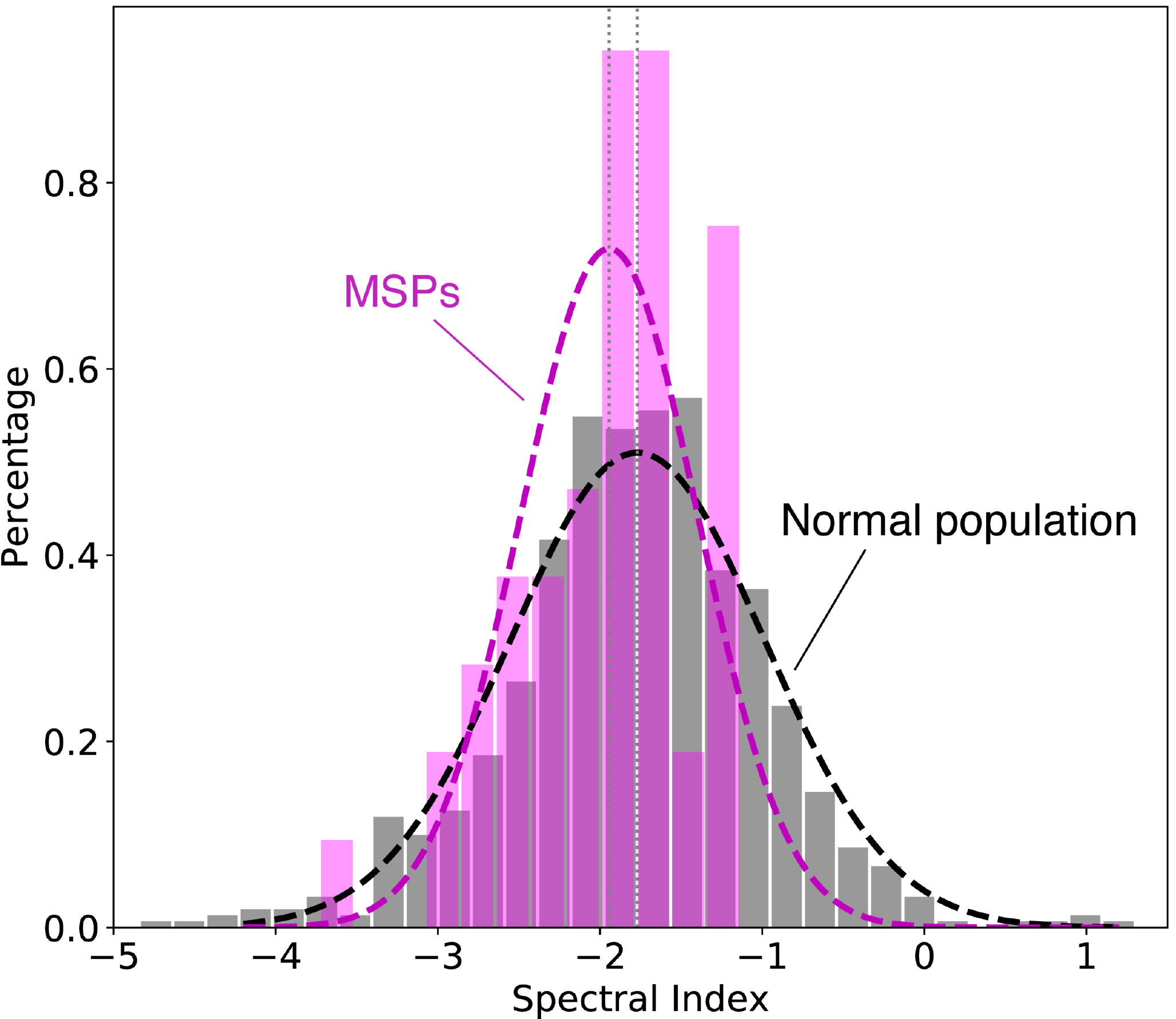}
    \caption{Histograms of spectral index distribution for the populations of MSPs ($P\leq\,$30$\,$ms) and normal population ($P>\,$30$\,$ms) from the database PSRCAT v1.62 \citep{man05}. A normal distribution is fit in each case, resulting in $\left<\alpha_{\rm MSP}\right>=-1.95\pm0.54$ for the MSP-only population, and $\left<\alpha\right>=-1.76\pm0.79$ for the rest.} \label{fig:avespec_hist}
\end{figure}

The putative Galactic centre population was derived using the Online Pulsar Catalog PSRCAT\footnote{\url{https://www.atnf.csiro.au/research/pulsar/psrcat/}} version 1.62 \citep{man05}. We extracted the pseudo-luminosity of those pulsars for which this information is available\footnote{The pseudo-luminosity ($L$) is defined as the flux density ($S$) multiplied by the square of the distance to the pulsar ($d$), $L=S\cdot d^2$ \citep[see e.g.][]{lorkra04}. As it contains information on the distance to the pulsar, it is used in our analysis instead of the flux density.}. In total, 2125 pulsars out of 2800 ($\sim$76\%) have pseudo-luminosity information in our PSRCAT dataset. However, the reported pseudo-luminosity is always for frequencies below those used in this survey; mainly at 400 ($\sim$8.5\%) and 1400$\,$MHz ($\sim$67\%). To asses any effect from possible spectral gigahertz-peaked turn-overs \citep[see e.g.][]{kiyak17}, we verified that the results did not vary when only the 1400$\,$MHz pseudo-luminosities were used or the 400 and 1400$\,$MHz information was included. We extrapolated the pseudo-luminosity to 120$\,$GHz (the central frequency of the 3+2~mm dataset) with the following method: If a pulsar had a measured spectral index available in PSRCAT, we used this spectral index for the extrapolation. If no spectral index was available for a pulsar, we drew a value from a normal distribution with mean value and standard deviation calculated from the 788 pulsars (i.e. $\sim$28\% of the total) in PSRCAT with known spectral indices, separating the MSP population\footnote{We define MSPs as pulsars with $P\leq\,$30$\,$ms. 51 out of 382 entries ($\sim$13\%) of MSPs in PSRCAT v1.62 include a spectral index. This spin period limit to separate populations is clearly simplistic \citep[see e.g.][]{kjlee12_GM}, but it suffices for the analysis presented here.} from the rest. The mean spectral index obtained for the MSP population is $\left<\alpha_{\rm MSP}\right>=-1.95\pm0.54$, and for the rest, it is $\left<\alpha\right>=-1.76\pm0.79$ \citep[see Fig.~\ref{fig:avespec_hist}, cf.][]{mar2000, jank18}. In both cases we assumed a single power law for the extrapolation, that is, $S_{\rm \nu}\propto \nu^\alpha$. The population simulation was repeated 5000 times to derive an error figure from the standard deviation. We note that the mean spectral indices derived for the two populations are statistically equivalent because they deviate by less than 1$\sigma$. However, our simulations of pulsar luminosities at millimetre wavelengths show that for similar average values, the standard deviation of the mean spectral index plays a fundamental role in the number of pulsars that may posses a sufficiently flat spectrum to be detectable\footnote{When a single mean spectral index is derived for the full population of Galactic pulsars, the result is $\left<\alpha\right>=-1.78\pm0.77$. The population coverage would not change significantly, but by using a single value, we could lose track of the potential lower luminosity at high radio frequencies that might in particular affect the MSP population.}. As a consequence, the MSP population, showing a smaller standard deviation in their spectral indices than the rest, tend to appear at a lower luminosity at higher radio frequencies. This property may affect the discovery potential of MSPs in our survey, adding to another effect that complicates the detection of fast MSPs due to uncorrected interstellar dispersion, as explained in Sec.~\ref{sec:resdis}.

Finally, we calculated the detection limit of the survey by converting our minimum detectable flux density into a pseudo-luminosity limit at the Galactic centre multiplying $S_{\rm min}$ by the square of the distance to the Galactic centre \citep[see e.g.][]{lorkra04}, $d_{\rm GC}=8.18\,\rm kpc$ \citep{gravc19},

\begin{equation}\label{eq:GClum}
    L_{\rm min}^{\rm GC} = S_{\rm min}\,d_{\rm GC}^2 = S_{\rm min}\,8.18^2.
\end{equation}

With the luminosity limit at the Galactic centre, we computed the number of detectable pulsars by the survey from the simulated population as those where the pseudo-luminosity at the observing frequency was higher than the limit (see Sec.~\ref{sec:resdis}).


\subsection{Mock pulsar injection}\label{sec:fakepsr}

The theoretical limit presented in Sec.~\ref{sec:sensitivity} assumes Gaussian noise from the instruments and does not take the potential effect of undesired interfering signals, red noise, of the filtering applied to reduce them, or the limitation of the searching algorithms to recover highly accelerated pulsars into account. To try to account for these factors and evaluate the sensitivity thresholds, we carried out an independent preliminary analysis for which we injected six synthetic pulsar signals into the real data for a large number of epochs and performed the searching process. We manually adjusted the flux density, $S$, of the injected mock signals until they reached the detection limit, defined as the flux density of the synthetic signals for which the pulsars are detected by the pipeline in good weather and long integration epochs, but missed in those epochs when the conditions are not optimal. To produce the mock pulsar signals, we used two codes: for the isolated pulsars, the \presto\  \texttt{injectpsr.py} routine, and for the pulsars orbiting companions, a custom version of \texttt{fake} from the software package \sigproc\footnote{\url{http://sigproc.sourceforge.net}}. The pulsed-averaged flux density of the synthetic pulsars was calculated by folding a noiseless time series with the pulsar signal and measuring $S$ as the integrated profile divided by the spin period. We took the masked fractions in the searched data into account, which mainly account for the gaps in between scans (see Sec.~\ref{sec:dataanalysis_dcp} and Table~\ref{tab:obs_table}). The masks have different sizes for different observations, and we used the median value of the fraction of filled gaps over the total time after concatenation (32\%) to calculate representative values of $S$. The properties of the injected pulsars and the empirically derived flux density values are summarised in Table~\ref{tab:mockpulsars}.

The mock data injection tests also allowed us to validate our data cleaning steps and to evaluate the limits of the algorithms in recovering highly accelerated signals in our data and optimise the significance thresholds both for \texttt{accelsearch} and the candidate-sifting step. In addition, these pre-processing tests, in particular when simulating tight orbits around black holes, served as a confirmation of the ability of the pipeline to recover the mock pulsars, and therefore similar real pulsar would they exist in the data\footnote{\presto, and in general, all typical pulsar searching software, is not confused when several, even many, pulsars co-exist in the data. Examples of this are the pulsar observations and searches in globular clusters \citep[see e.g.][]{clf2000} or observations of pulsars that are co-located on the sky and are covered within the beam size of a single observation.}. Finally, we established the need of a very high value for the parameter \texttt{zmax} of the \presto\  \texttt{accelsearch} for the most extreme systems \citep[see Sec.~\ref{sec:resdis}, also][]{eat21sub}.



\begin{table}
        \centering
        \caption{Parameters of the mock pulsar signals injected in the data during the tests to evaluate the sensitivity thresholds and find optimum parameters for the data analysis. The first three pulsars are isolated, and the last three orbit companions; the first orbits a stellar mass black hole and the other two orbit Sgr$\,$A*. In the binary systems the orbital inclination is 90 degrees and the orbital phase is that with a maximum acceleration. The columns indicate the intrinsic spin period ($P$), pulsar mass ($M_{\rm p}$), companion mass ($M_{\rm c}$), orbital period ($P_{\rm b}$), pulse-averaged flux density ($S$), and pulsar duty cycle ($\delta$).}
        \label{tab:mockpulsars}
        \begin{tabular}{S[table-format=3.2]ccccc} 
                \hline
        $P$                     &  $M_{\rm p}$          & $M_{\rm c}$       & $P_{\rm b}$        &       $S$     &       $\delta$\\
        {(ms)}              &  ($M_{\odot}$)    & ($M_{\odot}$) & (hr)                  &       (mJy)   &               \\
                \hline
        457.09      &  $-$                      & $-$               &  $-$          &   0.34    & 0.05   \\
        23.66       &  $-$                      & $-$               &  $-$          &   0.18    & 0.05   \\
        2.87        &  $-$                      & $-$               &  $-$          &   0.78    & 0.2   \\
        141.87  &  1.4                  & 10                &  30           &   1.09    & 0.1   \\
        55.22       &  1.4                      & $4.3\cdot10^6$    &  4320         &   0.82    & 0.1   \\
        1.92    &  1.4                  & $4.3\cdot10^6$    &  4320         &   0.86    & 0.2   \\ 
                \hline
        \end{tabular}
\end{table}

\section{Results and discussion}\label{sec:resdis}

\begin{figure*}
    \centering
    \includegraphics[width=\textwidth]{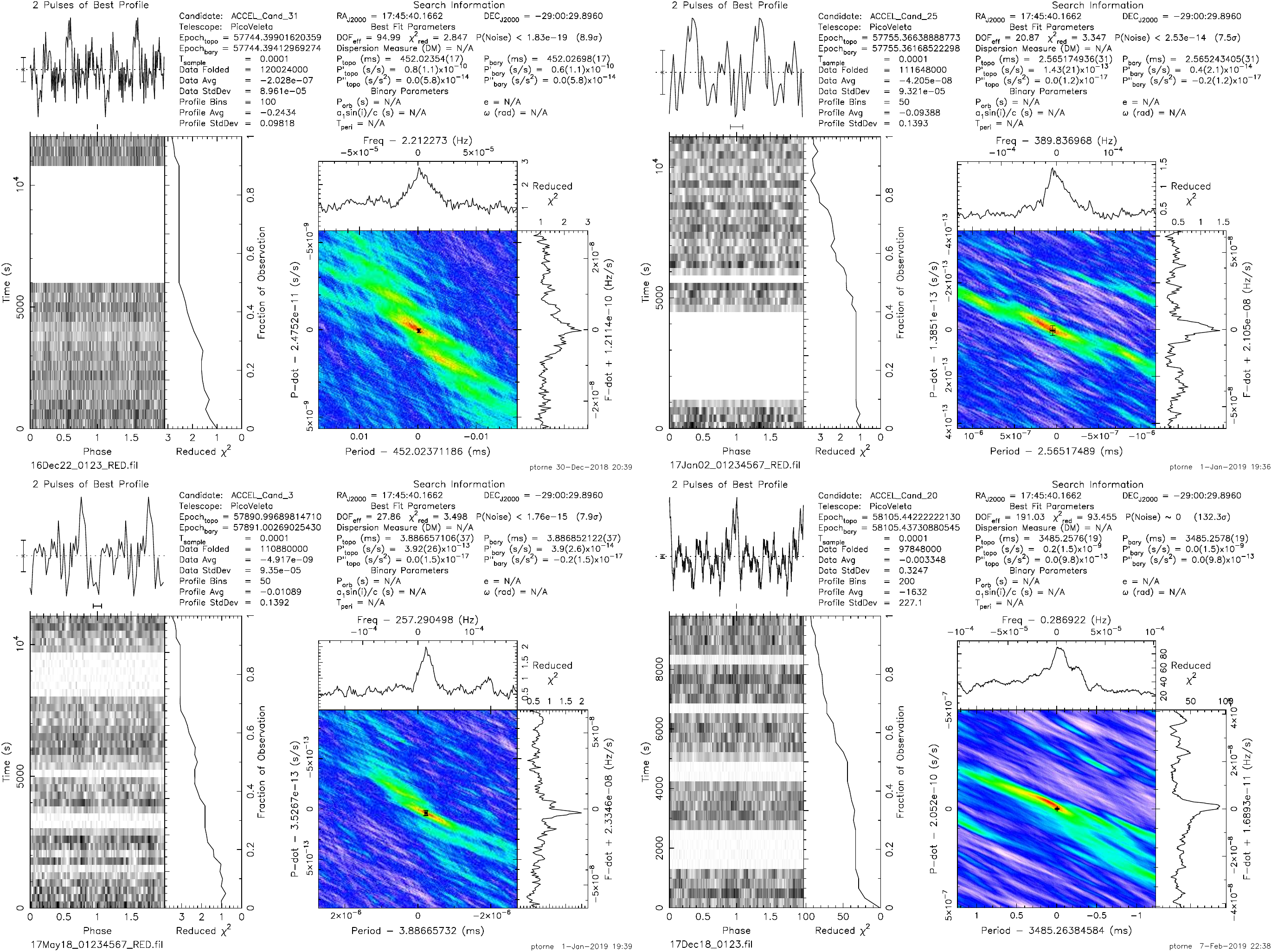}
    \caption{Four best candidates from the IRAM~30m pulsar survey as evaluated from a combination of detection significance and the characteristics of the candidate after folding. The main properties on which our selection is based are persistence of the signal in time, significance in the period-period derivative panel, $\sigma$ value, and the profile shape. A list with the properties of these four and the other selected best candidates is given in Table~\ref{tab:bestcands}. The upper left panel shows a candidate with a close period to one of the synthetic injected pulsars, but we remark that the candidate arises from data with no mock pulsar injected. The bottom right panel candidate has a relatively close period to PSR$\,$J1745$-$2900, but it seems to be harmonically unrelated. A description of the different panels of the plots is presented in the caption of Fig.~\ref{fig:J1745_rednoise}.}
    \label{fig:bestcands}
\end{figure*}

The search of the presented dataset produced a total of 5431 pulsar candidates, yielding 10862 plots that were individually reviewed\footnote{Each candidate produces two plots (see Sec.~\ref{sec:dataanalysis_ps}).}. The most prominent candidate repeatedly is the Galactic centre magnetar PSR~J1745$-$2900, which is detected at a high significance in most epochs. In addition to PSR~J1745$-$2900, a number of potential pulsar candidates were identified, but most showed a relatively low significance, and all appeared only once. Without a high-significance and consistently repeated detection of a given candidate periodicity, we considered that no new pulsars were distinctly discovered in this survey. Because a detection of any of these candidates at lower frequencies, where they could be brighter and where information of the dispersion measure can be obtained, could work as a confirmation of a potential real new pulsar, we add the properties of our 20 best candidates in Table~\ref{tab:bestcands} and show the 4 best candidate plots in Figure~\ref{fig:bestcands}. These candidates were chosen from a combination of detection significance and their pulsar-like characteristics in the candidate plots after folding.

\begin{table}
        \centering
        \caption{Parameters of the 20 best candidates from the IRAM~30m Galactic centre survey excluding the Galactic centre magnetar PSR~J1745$-$2900. Columns indicate the barycentre spin period of the candidate ($P_{\rm bary}$), the significance of the signal as given by \presto\  \texttt{sifting.py} ($\sigma$), the acceleration estimated by \texttt{accelsearch} (accel.), the epoch of the observation, and the frequency band (Dataset) that produced the candidate. An asterisk after the frequency band indicates the candidates shown in Fig.~\ref{fig:bestcands}.}
        \label{tab:bestcands}
        \begin{tabular}{S[table-format=3.4]S[table-format=2.1]S[table-format=-2.2]cc} 
                \hline
        $P_{\rm bary}$  &  $\sigma$         & {accel.}      & Epoch         & Dataset       \\
        {(ms)}              &               & {(ms$^{-2}$)} &               &             \\
                \hline
        452.0269            &  9.1                  & 0.00    & 2016 Dec 22     &   3$\,$mm*   \\
        2354.2995       &  10.5         & 0.00    & 2016 Dec 22     &   2$\,$mm \\
        9.3334       &   9.0         &  0.02   & 2017 Jan 02     &   3+2$\,$mm \\
        2.5652      &   6.4         &  0.00  & 2017 Jan 02     &   3+2$\,$mm* \\ 
        5746.5523       &   4.1        & -33.50      &   2017 Jan 23     & 3+2$\,$mm \\
        29.0021      & 3.2           &  0.00   & 2017 Feb 14      & 3$\,$mm   \\
        23.9634      &   3.6        &  -9.90         & 2017 Mar 01     &   3+2$\,$mm  \\
        12.1925     & 3.7           &   40.50        & 2017 Mar 01     & 3+2$\,$mm \\
        5.7427      &   6.2        &  0.00         & 2017 Mar 10       & 2$\,$mm \\
        3.8184     &   11.8        &  0.00      & 2017 Apr 18   &  3$\,$mm   \\
        3.8868     &   7.1     &  0.00     &   2017 May 18 & 3+2$\,$mm*   \\
        2.0782       & 5.9   & 0.00      &   2017 May 24 &     3$\,$mm   \\
        3.3240     &   6.1     &   -0.24 &   2017 Jun 05 & 3+2$\,$mm \\
        2.3460      &   7.0     &    0.05       & 2017 Nov 20   & 3$\,$mm   \\
        8.4191      &   7.5     &  0.04         & 2017 Nov 20    & 2$\,$mm \\
        3485.2578        & 6.6   &   0.00        &   2017 Dec 18     & 3$\,$mm*   \\   
        9.7925      &   2.2     & 0.00  & 2017 Dec 18       & 3$\,$mm   \\
        22.8158     &   2.8     & 0.00              & 2018 Jan 15       &    3$\,$mm   \\
        14.3172     &   8.5     & 0.02         & 2018 Jan 15      &   3$\,$mm   \\ 
        2.1344         &   5.6     &   0.00       & 2018 May 15       &  2$\,$mm \\
\hline
        \end{tabular}
\end{table}

\begin{figure}
    \centering
    \includegraphics[width=\linewidth]{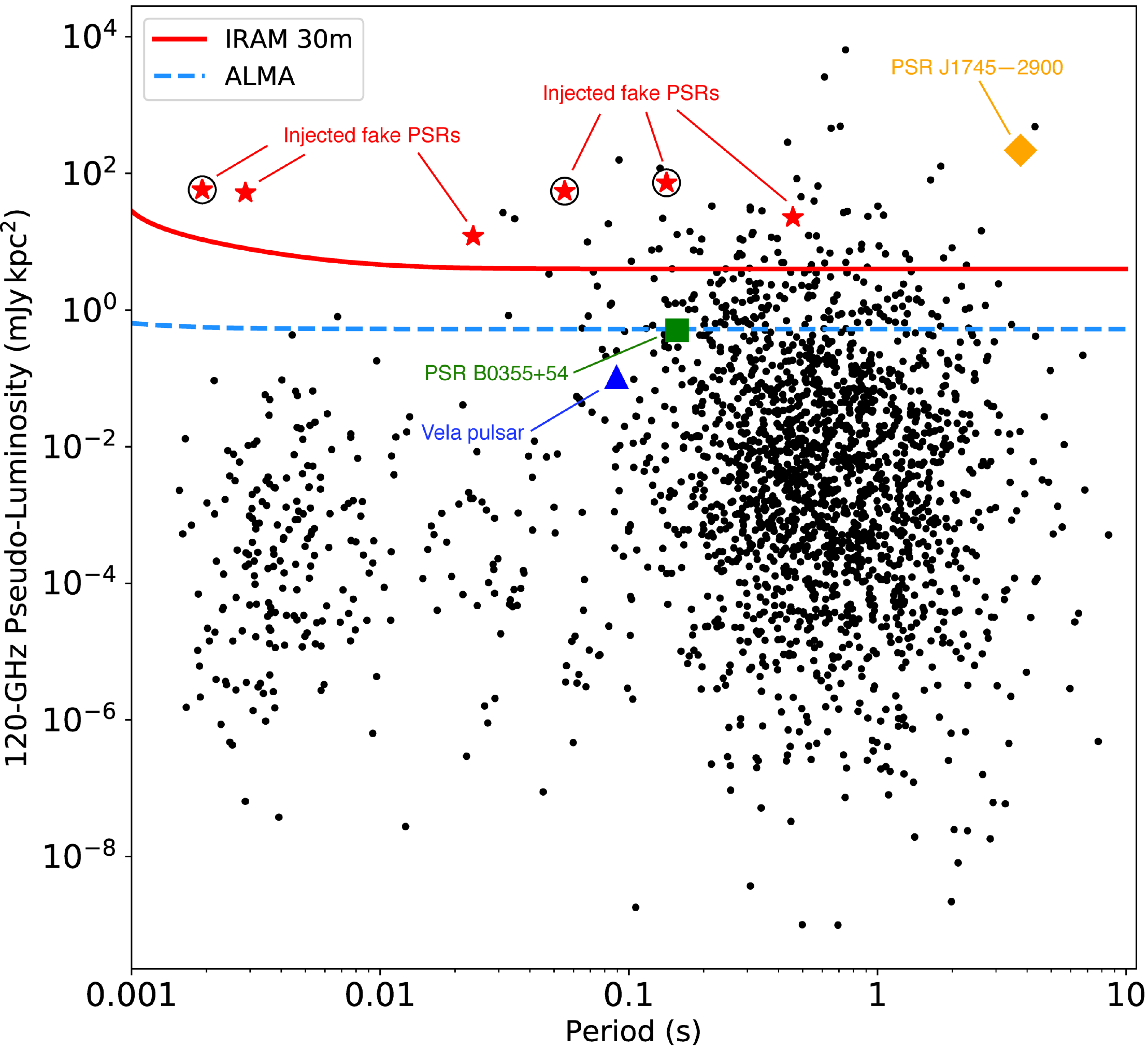}
    \caption{Sensitivity to pulsars in the Galactic centre by this survey (continuous red line) and a hypothetical survey with ALMA at similar frequencies (dashed light blue line). The black dots represent the pseudo-luminosities for the known population of pulsars extrapolated to 120$\,$GHz (see Sec.~\ref{sec:sensitivity} for details), simulating a population of pulsars in the Galactic centre. The dots above each corresponding threshold line are pulsars that may theoretically be detected by the surveys. The sensitivity lines are calculated assuming a duty cycle for the minimum detectable flux density of $\delta=0.1$. We assumed a hyper-strong scattering towards the Galactic centre following the NE2001 model, $\tau_s \approx 2000 \, \nu_{\rm GHz}^{-4}\,$s \citep{cl2002} to consider a worst-case scenario. The effect of scattering is negligible (see text). The dominant interstellar effect is the uncorrected dispersion smearing for the IRAM~30m data, which is apparent at the very end of the short spin periods as a raising limit. The dispersion smearing effect is not affecting the ALMA limit because we consider that the ALMA data will enable the correction for dispersion. The red stars indicate the synthetic pulsar signals injected for the sensitivity tests. The three circles around the red stars with $P\simeq\,$1.92, 55.22, and 141.87$\,$ms mark the three pulsars orbiting black holes that were injected to test the sensitivity to highly accelerated pulsar systems. The orange diamond shows the average pseudo-luminosity of the Galactic centre magnetar PSR$\,$J1745$-$2900, and the blue triangle and green square mark the 120$\,$GHz pseudo-luminosities of the Vela pulsar (PSR$\,$B0833$-$45) and PSR$\,$B0355+54, respectively.}  \label{fig:sensitivity_plot}
\end{figure}

When we assume that pulsars exist in the Galactic centre, two main reasons explain the lack of new discoveries. On one hand, the decreased flux density of pulsars at short millimetre wavelengths due to their typically steep spectral index. On the other hand, the current sensitivity of the IRAM~30m telescope, which even though it is very high for a facility operating in this frequency regime, is relatively low when compared to 100 m class telescopes and may be currently insufficient to detect the weak pulsar signals. Our analysis of the survey sensitivity following Sec.~\ref{sec:dataanalysis} results in a representative minimum detectable flux density of $S_{\rm min}^{\rm 30m}=59\,\mu$Jy. The same limit expressed in terms of pseudo-luminosity at the distance of the Galactic centre (following Eq.~\ref{eq:GClum}) is ${L_{\rm min}^{\rm GC}}\simeq 3.95\, \rm{mJy\,kpc^2}$. For a putative Galactic centre pulsar population resembling that of the Milky Way, only about 3.5$\pm$0.3\% of the pulsars could be detected with this limit. The error figure of 0.3\% shows the 1$\sigma$ statistical standard deviation related to the population simulation (see Sec.~\ref{sec:sensitivity}). However, when mock pulsar signals are used to evaluate a sensitivity limit, the result is even lower. The analysis shows that our detection threshold can sometimes be higher by up to a factor $\sim$10 than the theoretical one (see the $S$ column of Table~\ref{tab:mockpulsars} and the position of the synthetic pulsars with respect to the theoretical limit in Fig.~\ref{fig:sensitivity_plot}). With these higher thresholds, our survey would be sensitive to only about 2\% of isolated pulsars and about only 1\% of those in highly accelerated binary systems. We furthermore remark that our coverage to MSPs is close to zero. The latter fact arises from the simulation showing that under the assumptions on spectral index we made, the number of MSPs possessing a flat spectrum that therefore are bright enough to be detected at our observing frequencies is much lower than for the remaining pulsar population.

We relate the decreased sensitivity shown by the mock pulsar analysis to the characteristics of the noise from the EMIR receiver, and for the binary systems, to an additional component from the limitation of the algorithms in recovering pulsar signals of highly accelerated systems. For example, for the synthetic binary in a 30 hr orbit of a 10 $M_{\odot}$ black hole, even with a \texttt{zmax}=1200, \texttt{accelsearch} can only recover eight harmonics in our datasets. For the fast-spinning MSP ($P=$1.92$\,$ms) in a six-month orbit of Sgr$\,$A* only one harmonic is recovered. We note, nevertheless, that the limitations of \texttt{accelsearch} occur only for extremely accelerated systems and are largely augmented by our long total integration times after concatenating the scans filling the gaps in between them (see Sec.~\ref{sec:dataanalysis}). For a more detailed discussion of the limits for accelerated pulsar recovery with \presto, including effects such as jerk or eccentricity, see \citet{eat21sub}, \citet{liu21sub}.

Furthermore, the effect of the duty cycle of the potential pulsars to be detected is non-negligible (see Eq.~\ref{eq:radiometer}). For instance, a shorter duty cycle of $\delta=0.05$ would increase the theoretical percentage of average detectable population to $\sim\,$4.4\%, and a duty cycle $\delta=0.4$ would lower it to $\sim\,$1.9\%. For comparison, our simulated survey with ALMA (for $\delta=0.1$) would theoretically cover about 9.6\% of the population (11.1\% if the derived sensitivity from QA2 is used), showing one path towards higher-sensitivity millimetre-wavelength surveys in the future. Figure~\ref{fig:sensitivity_plot} illustrates the sensitivity thresholds over the simulated Galactic centre population, including our synthetic injected pulsars, and three real pulsars for reference.

We remark that we may have an additional component of sensitivity loss to the MSP population. 
Firstly, MSPs in binary systems are more challenging to detect with the applied searching algorithms because the drifts in frequency space due to Doppler effects are larger \citep[and therefore are more difficult to correct, see][]{eat21sub} than for slow pulsars. Secondly, our inability to correct for interstellar dispersion with a continuum backend could translate into a lower sensitivity to fast-spinning MSPs (with periods $P\lesssim3\,$ms) due to pulse smearing. This last issue is one reason why three separated datasets were searched (3, 2, and 3+2$\,$mm, see Sec.~\ref{sec:dataanalysis_dcp}). The dispersion delay for the 3~mm-only data (between 104 to 84$\,$GHz) for a dispersion measure (DM), DM=2000~pc$\,$cm$^{-3}$, is $\Delta t_{\rm DM}\approx\,$0.37$\,$ms. For 2~mm-only (156 to 136$\,$GHz), the dispersion delay would be $\Delta t_{\rm DM}\approx\,$0.11$\,$ms. In the case of 3+2~mm (156 to 84$\,$GHz), a DM=2000~pc$\,$cm$^{-3}$ would produce a frequency-dependent delay of $\Delta t_{\rm DM}\approx\,$0.83$\,$ms. In this latest case, the dispersion is appreciably larger than the sampling interval and can therefore smear a sufficiently narrow pulse. We graphically show in Fig.~\ref{fig:sensitivity_plot} the corresponding potential decreased sensitivity to MSPs by including the effect of uncorrected dispersive smearing in the IRAM~30m data through Eqs.~\ref{eq:weff} and \ref{eq:radiometer}, assuming a DM=2000~pc$\,$cm$^{-3}$. 
The DM=2000~pc$\,$cm$^{-3}$ was chosen based on the DM measured for the closest pulsar to Sgr$\,$A* known today: PSR~J1745$-$2900 with DM=1778 pc$\,$cm$^{-3}$ \citep{eat13}. Even so, the DM closer to Sgr$\,$A* could be even higher for example by the contribution of a denser gas or even an accretion disk \citep{cl2002}, which would reduce our ability to detect short-period MSPs in this survey even further.

To the previously outlined challenges of detecting fast-spinning MSPs, we add that the probability is high that due to the very high stellar densities of the Galactic centre (and thus a much higher probability of stellar interactions), a good portion of the pulsars populating the region could be MSPs \citep[e.g.][]{whar12, macq15}. A scenario in which the Galactic centre pulsar population is dominated by MSPs could therefore explain the lack of discoveries in our survey. Interestingly, the majority of the best candidates yielded by the IRAM~30m survey are MSPs, with $P<$10$\,$ms. The limitations discussed above make us cautious to consider these MSP candidates as real, but the results might at the same time be indicative of an MSP-dominant population in the Galactic centre.

In addition to optimising the potential effect of uncorrected interstellar dispersion, there is a second reason for searching the three frequency datasets separately. It relates to the EMIR locally generated periodic signals that are more prominent in the 3$\,$mm band than in the 2$\,$mm band\footnote{The origin of these signals is under investigation, but the first hypothesis points to possible oscillations in the bias circuits of the first mixers of the receiver.}. As a result, when we combined the 3 and 2$\,$mm bands, we increased the theoretical sensitivity by increasing the bandwidth, but we added these undesired signals as well. On the other hand, the 2$\,$mm band of EMIR is more stable and produces cleaner noise, so that the Fourier spectrum is also cleaner. Despite the effects of the larger and somewhat less-clean noise at 3$\,$mm, we verified during the mock injected signal analysis and with the blind detections of PSR$\,$J1745$-$2900 that the combination of the 3 and 2~mm bands can increase the detection capability for weak pulsars with relatively flat spectra. In summary, the 3$\,$mm dataset is potentially better suited to detect pulsars with not-so-flat spectral indices. The 2$\,$mm band shows a better-behaved noise, but it is not optimal because the frequency is higher and pulsars tend to be dimmer. The 3+2$\,$mm data are finally the most sensitive with their larger combined bandwidth, but shows the undesired stronger interference from the combined frequency bands. To minimise the potential adverse effects and maximise the chances of detecting pulsars in our full dataset, we searched the three datasets individually.

Some of the locally generated signals are periodic and polluted certain regions of the Fourier series. 
The \texttt{rednoise} filter of \presto$\,$ reduced them in part, but their intensity was sometimes so strong that we considered that parts of the Fourier series cannot be used to detect pulsars. 
Consequently, we might lose pulsars even with luminosities above our detection thresholds. This would nonetheless only occur if a pulsar rotational frequency lies exactly within a bin in the power spectrum that is affected by the local periodic signals. We estimate that $\sim$3.5\% on average of the Fourier bins are affected. Although this is not too significant, it is an additional cause for potential sensitivity loss.


A final challenge to mention is the effect of the red noise. It has been demonstrated that pulsar surveys lose sensitivity to long-period pulsars when red noise is present in the data \citep{laz15}. Our IRAM~30m data, and generally all pulsar observations at very high radio frequencies where variable opacity from the atmosphere contributes significantly to the system noise, show a considerable amount of red noise. The running-fit filter prior to searching has been developed to be effective at reducing this red noise while at the same time not affecting significantly potential pulsar signals. This has been verified by optimising detections of PSR$\,$J1745$-$2900, with a spin period $P\simeq$3.76$\,$s. We therefore successfully reduced the effect of the red noise in our survey, but we cannot eliminate it completely. We thus consider our sensitivity to very long-period pulsars ($P\gtrsim$4$\,$s) reduced further than the limits presented above.


Despite the limitations and challenges discussed earlier, the method we followed here is capable of detecting new pulsars. This is demonstrated by previous successful detections of pulsars with the EMIR receiver at the IRAM~30m telescope that covered the same frequency ranges as we used in this survey \citep{tor15,tor17,tor20_ATel}, the injection of simulated pulsar signals correctly detected in our pipeline, and the consistent blind detection of the magnetar PSR~J1745$-$2900 by our searching algorithm. To place our detection thresholds further into context, PSR~J1745$-$2900 belongs to the 0.5\% most luminous pulsars at 120~GHz \citep{tor15,tor17}, and PSR B0355+54 (one of the brightest pulsars with one of the flattest spectra known) belongs to the 10\% most luminous population at 120~GHz \citep[][Torne et al. \emph{in prep.}]{morr97}. The Vela pulsar (PSR$\,$B0833$-$45) is very bright at 1.4~GHz, but its spectral index of $\alpha\simeq\,-$1.5 \citep[taken for the component with the flatter spectrum,][]{keith2011} places its luminosity at $\sim\,$100$\,$GHz \citep{liu19} at a level below even the ALMA threshold. These examples show that we only reach the highest-luminosity population in this survey, and at the same time, that we are highly biased to flat-spectrum pulsars when we search in the short-millimetre regime. It would therefore be reasonable to believe that there may still be pulsars in the Galactic centre that have not yet been detected simply because they are dim at short-millimetre wavelengths and are beyond the reach of our current sensitivity.



The survey presented here has two remarkable advantages over previous surveys carried out at typical pulsar-observing wavelengths in the centimetre regime. In the first place, the scattering is negligible. Eqs.~\ref{eq:weff} and \ref{eq:radiometer} show how the temporal scattering would normally affect the sensitivity to detect pulsars, although the effect can be more complex, for instance affecting the periodicity-search algorithms \citep{kra2000, mq10}. Even if we consider a worst-case scenario and assume a hyper-strong scattering to the Galactic centre with a scattering time of 2000$\,$s at 1$\,$GHz \citep[NE2001 model,][]{cl2002}, the scattering in our lowest frequency range would be $\tau_s \approx 2000 \cdot 86^{-4} = 36\, \rm{\mu s}$ \citep[see e.g.][]{lorkra04}. This scattering is negligible even for the fastest-spinning pulsars known. For comparison, in the same situation, a survey at an observing frequency of 2$\,$GHz ($\uplambda=$15$\,$cm) would be exposed to a scattering time of $\tau_s = 125\,$s, and even at 5$\,$GHz ($\uplambda=$6$\,$cm), this would be $\tau_s = 3.2\,$s. These long scattering times would disrupt the detection of pulsations at centimetre wavelengths. Even in the case of much lower scattering towards the Galactic centre, the population of fast-spinning MSPs may still be beyond the reach of surveys below $\sim$5$\,$GHz \citep{spi14}. 
The second advantage of the survey is the very low dispersion at millimetre wavelengths. This means that this parameter space does not require an intensive search. This situation allows us to save computing time that can be spent surveying the acceleration parameter space. 
Thus, this first short-millimetre-wavelength survey with the IRAM~30m telescope covers the parameter space of highly scattered and highly accelerated pulsars in the Galactic centre region for the first time \citep[cf. e.g.][]{john06, den09, mq10, bat11, whar12p1, eat13b, siem13}, and it is virtually unbiased to population coverage regardless of the scattering characteristics in this direction.

We note that the small effect of interstellar dispersion that allows us to search using a continuum backend and to expand our search space in acceleration becomes a great disadvantage in terms of discriminating interference or local signals from celestial ones. On one hand, the dispersion correction in a pulsar-searching algorithm at low radio frequencies smears local broadband interference, reducing its effect in the periodicity search, especially at high DM values such as are expected in the Galactic centre. At short millimetre wavelengths such as we used in the presented survey, we greatly limit this ability to suppress local broadband signals, and the local interference enters the search algorithm at almost full strength. This has two main consequences. First, as discussed earlier, the periodic interference may contaminate specific parts of the Fourier series, with the potential to hinder the detection of a pulsar with an equal or close spin period to the period of the interfering signal. On the other hand, the interference will increase the number of candidates produced by the search, sometimes very significantly, with the consequent additional difficulty in the candidate sifting and review steps. Related to this issue, the lack of spectral information of the pulsar candidates in our search (from using a continuum backend) affects our capability to discern broad- and narrow-band signals. It was therefore not possible to identify pulsar-resembling candidates produced by narrow-band locally generated signals.


Finally, it is remarkable that this survey covers a long time span, with repeated observations. This is an advantage compared to single-epoch or few-epoch surveys, and makes the search robust against time-varying factors that can prevent the detection of pulsars. One of the factors is weather (atmospheric opacity and turbulence), which can affect the sensitivity of short-millimetre observations greatly if, when only a few epochs are observed, the weather is not optimal. Other important factors whose effects are diminished by the repeated observations are the potential geodetic precession of pulsars in orbit with massive objects due to curved space-time, which can move the radio beams in and out of our line of sight \citep[e.g.][]{des19}, or eclipses that can temporarily hide the pulsar radio emission \citep[e.g.][]{lyne1990}. Lastly, the repeated observations at different epochs increase the opportunities of observing an extreme binary system in parts of the orbit in which the acceleration effects are less detrimental for the signal recovery in the searching algorithm, for example where the jerk is low \citep[see e.g.,][]{eat21sub}.

When we look into the future and focus on short-millimetre wavelengths, one logical way to improve the surveys is by increasing the sensitivity by using larger telescopes such as the Large Millimeter Telescope (LMT), the NOrthern Extended Millimetre Array (NOEMA)\footnote{NOEMA is developing a phasing mode with the potential to enable the use of the interferometer as a $\sim$52-m equivalent dish with pulsar observing capabilities.},
or ALMA. From the instrumental side, other methods for increasing the sensitivity include the extension of the instantaneous bandwidth, the use of different receiver technology such as Kinetic Inductance Detectors \citep{tor20}, and the use of fast spectrometers to enable dedispersion and obtaining the spectral information of the candidates. Other improvements may arise from the searching algorithms, for example by applying the fast-folding algorithm \citep[FFA, see e.g.][]{cam17} or extending the acceleration and jerk parameter spaces surveyed. In particular at very high radio frequencies, a more advanced data cleaning applying smart filters to reduce the undesired potential interfering signals and the red noise while not affecting potential pulsar signals can also help to increase our sensitivity. Surveying using the frequency range of $\nu\simeq$30$-$50$\,$GHz ($\uplambda\simeq$10$-$6$\,$mm) would also be compelling because the scattering will still be largely reduced but pulsars are more luminous on average. To take advantage of long observing campaigns, searching in an incoherent or even a coherent stack of a number of epochs promises to significantly improve the sensitivity \citep{eat13b, pan16, lentati18}. Finally, another procedure to try to find new pulsars and other signals such as those from rotating radio transients \citep[RRATs, see e.g.][]{keaneMc_2011} or fast radio bursts \citep[FRBs, see e.g.][]{corsha19} is through the search for single pulsations or transient emission. We did not apply this technique in our survey because in our experience, an insufficient dispersion effect and frequency information to filter local interfering transient signals prevented an efficient transient-like candidate classification.

In spite of challenges such as the large distance, potential strong scattering, unknown population, and sensitivity limitations, the undeniable importance of finding pulsars in the Galactic centre and in particular, those orbiting Sgr$\,$A*, justify additional and continued efforts to survey this region. Surveys in the short-millimetre regime are novel and need to be further improved, but they offer a new tool in our quest to uncover the hidden population of pulsars in the centre of the Milky Way and behind extremely scattered regions.

\section{Summary and conclusions}\label{sec:sumcon}


We presented a targeted pulsar survey of the Galactic centre that for the first time used short-millimetre wavelengths. After the analysis of a total of 62.2 hours of observations in 28 different epochs in a period spanning almost 1.5 years, no clear detections of new pulsars were found. The non-detections can be explained because the sensitivity of the observations is limited. It was estimated that only about 2 and 1\% of a hypothetical Galactic centre population located close to Sgr$\,$A* would be detected for isolated and highly accelerated binary pulsars, respectively. For the particular case of Galactic centre MSPs, the current sensitivity appears insufficient to expect detections at short-millimetre wavelengths under our assumptions for population simulations. 

The main reasons for the limited sensitivity to pulsars in the Galactic centre are a combination of the use of very high observing frequencies to overcome the scattering with the typical steep spectral index of pulsars in the radio band, making pulsars extremely dim at millimetre wavelengths, the large distance to the centre of the galaxy, and the relatively low sensitivity of the IRAM~30m telescope mainly due to its dish size (when compared to 100$\,$m class telescopes that are generally used for pulsar science). We also found an additional loss of sensitivity from other factors. Firstly, from a considerable amount of red noise in the data, typical from slow atmospheric opacity variations during the observations. The excess of red noise mainly affects the detection of long-period pulsars. This red noise effect was reduced by our filtering steps prior to the search by using a running-fit filter. Nonetheless, the detrimental effect of red noise cannot be fully eliminated. Second, periodic locally generated signals generated inside the EMIR receiver pollute parts of the Fourier series, which diminishes our detection capability for certain particular periodicities. Although this only affects a small percentage of our Fourier series, we successfully used the \presto\  \texttt{rednoise} filter to significantly reduce these undesired signals, but for certain regions of the Fourier spectrum, we may still lose real pulsars if the locally generated signals have periods equal to the potential pulsar spin period. Finally, we were unable to correct for dispersive smearing by using a continuum backend. This had two main drawbacks: an additional loss in sensitivity to fast-spinning MSPs, and the impossibility of distinguishing locally generated periodic or interfering signals mimicking pulsars from potential real pulsar signals due to the lack of DM measurement. In addition, we were unable to differentiate pulsar candidates produced from narrow- and broad-band signals, making the candidate classification less effective. For this last obstacle, future observations with a very high frequency and temporal resolution, or at a lower observing frequency, may be needed to fully confirm a candidate by the detection of a dispersion measure different from zero and the confirmation of the broad-band nature of the emission as expected for radio pulsars.

In contrast, the principal advantages of searching for pulsars using short-millimetre wavelengths are a reduction of the scattering to negligible levels, being effectively unbiased in population coverage, and the possibility of searching a much larger acceleration and jerk parameter space for a given computing time in exchange for the very small (or negligible) space of dispersion measures to be blindly searched. Despite the sensitivity and the challenging data reduction, this pulsar survey with the IRAM~30m telescope is the first to cover a parameter space that is basically unaffected by temporal scattering and thus has the potential of detecting fast-spinning MSPs in the Galactic centre if the scattering times are long in this direction.
The survey results reveal that no fundamental problem prevents us from carrying out pulsar surveys using a standard pulsar-searching software at these very high radio frequencies, and therefore motivates other similar surveys, conceivably with more sensitive instruments.





The interest of searching the IRAM~30m data is proven by the detections in our blind-search pipeline of the magnetar PSR$\,$J1745$-$2900 and a number of mock pulsar signals, including a pulsar-stellar black hole system and two pulsars in close orbit with Sgr$\,$A*. Our tests show that the survey was capable of detecting new pulsars if they were sufficiently bright, and the non-detections rule out that other bright radio magnetars similar to PSR$\,$J1745$-$2900  or highly luminous pulsars in the frequency ranges covered and with beams pointing towards us exist in the inner region of the Galactic centre, at least during the time span covered by the survey.


Similar pulsar surveys with improved sensitivity and involving larger telescopes, such as the LMT, NOEMA, and ALMA are encouraged. In the future, the use of new instrumentation with better performance in pulsar observations at millimetre observatories will help us to further constrain the characteristics of the potential pulsar population living inside the inner parsec of our galaxy.

\begin{acknowledgements}
The authors thank the anonymous referee and the language editor for the quick review of the manuscript. PT wishes to thank Scott Ransom for his help to clarify details on the \presto$\;\,$routines, Ciriaco Goddi and Sergio Dzib for useful comments on the draft, and the staff of the IRAM~30m telescope for their great support during the observations.
This work is based on observations carried out with the IRAM 30m telescope. IRAM is supported by INSU/CNRS (France), MPG (Germany) and IGN (Spain). Financial support by the European Research Council for the ERC SynergyGrant BlackHoleCam (ERC-2013-SyG, GrantAgreement no. 610058) is gratefully acknowledged. RE is supported by a ``FAST Fellowship'' under the ``Cultivation Project for FAST Scientific Payoff and Research Achievement of the Center for Astronomical Mega-Science, Chinese Academy of Sciences (CAMS-CAS)''. This research has made use of NASA's Astrophysics Data System.
\end{acknowledgements}

%
\bibliographystyle{aa} 
\bibliography{Torne_IRAM_GCsurvey} 
%


\end{document}